\documentclass[aps,prb,twocolumn,10pt,amsmath,amssymb,bibnotes,superscriptaddress,longbibliography]{revtex4-1}

\usepackage{graphicx, braket}
\usepackage{siunitx}
\usepackage[colorlinks,urlcolor=blue,citecolor=blue,linkcolor=blue]{hyperref}
\newcommand{\be}{\begin{equation}}
\newcommand{\ee}{\end{equation}}
\newcommand{\bea}{\begin{eqnarray}}
\newcommand{\eea}{\end{eqnarray}}
\newcommand{\cmmnt}[1]{\ignorespaces}

\usepackage[normalem]{ulem}

\begin{document}

\title{Resonant tunneling in graphene-ferroelectric-graphene junctions}




\author{David Koprivica}
    \affiliation{Raymond and Beverly Sackler School of Physics and Astronomy, Tel-Aviv University, IL-69978 Tel Aviv, Israel}
    \author{Eran Sela}
    \affiliation{Raymond and Beverly Sackler School of Physics and Astronomy, Tel-Aviv University, IL-69978 Tel Aviv, Israel}
\date{\today}

\begin{abstract}
We study tunnel junctions consisting of a two-dimensional ferroelectric material sandwiched between graphene electrodes. We formulate a theory for the interplay of the polarization and induced free charges in such devices, taking into account quantum capacitance effects. We predict a gate sensitive voltage difference across the polar domains, which can be measured using electrostatic force microscopy. Incorporating this electrostatic theory in the tunneling current-voltage characteristics, we identify a resonance peak associated with aligned Dirac cones as a highly sensitive probe of the polarization. This opens the way for device applications with few atom-thick polar layers acting as readable ultra-high-density memory.
\end{abstract}

\maketitle

\section{Introduction}
\label{se:intro}
Recent breakthroughs in two-dimensional (2D) van der Waals materials  led to the experimental realization of a new form of ferroelectricity (FE)~\cite{li2017binary,yasuda2021stacking,ViznerStern2021,weston2022interfacial,de2021direct,Swarup2022}. 
This newly discovered interfacial ferroelectricity results from polar stacking configurations that break inversion symmetry, such as in AB or BA stacked  hexagonal boron nitride (h-BN) bilayer~\cite{yasuda2021stacking,ViznerStern2021} or other 2D materials~\cite{weston2022interfacial,de2021direct,Swarup2022}.  Remarkably, the resulting polarization can be flipped~\cite{ViznerStern2021} by a relative sliding of the layers by a single atomic distance. 

Understanding the response to electric fields and the nature of ferroelectricity~\cite{wu2021sliding,zhong2021sliding} in these experiments requires further study and a possible distinction of these various systems~\cite{PhysRevB.105.235445}. In the bulk of twisted interfaces there is a Moir\'e-triangular lattice of alternating AB and BA regions which expand or shrink at finite field~\cite{de2021direct}. Yet, \emph{in average} over many Moir\'e unit cells there is no remnant polarization at zero field~\cite{Bennett2022}. 
On the other hand, in Refs.~\onlinecite{yasuda2021stacking,ViznerStern2021} hysteresis was seen, and specifically in Ref.~\onlinecite{ViznerStern2021}, the  flipping of a polar domain was attributed to the sliding of a domain wall near the edge of the bilayer, rather than in the bulk~\cite{PhysRevB.105.235445}. 
Leaving the precise sliding FE mechanism aside, here we shall be concerned only with uniform AB and BA polar interfaces.

Ferroelectric tunnel junctions, consisting of a thin FE material sandwiched between two electrodes, permit reading the  polarization via 
tunnelling electroresistance (TER)~\cite{pantel2010electroresistance,garcia2014ferroelectric,tao2016ferroelectricity,kang2020giant,yan2022giant}. What is the fate of interfacial polarization within a tunnel junction? 



In this work, we theoretically study
2D tunnel junctions consisting of uniform AB or BA polar interfaces sandwiched by graphene electrodes, as shown in Fig.~1(a). Assuming a given value for the bare 
polarization $V_{P}^{(0)}$, which, in principle, can be determined by \textit{ab initio} methods, we focus on the  interplay of  the polarization and screening charges forming on the graphene electrodes. 
Related questions were recently addressed within  density functional theory (DFT)~\cite{yang2021sliding}, where it was found that whereas metallic electrodes significantly affect the polarization of a bilayer h-BN, adding graphene spacers between the FE and metals restores the polarization and results in a significant TER.  Here, we provide a phenomenological electrostatic model.
We find that when one of the electrode's Fermi level is tuned to the Dirac point, the electronic equilibration is dominated by quantum capacitance, and then a polar domain associated local voltage can be measured across the device. 

We focus on basic mechanisms allowing to read out the 
polarization direction and magnitude from the  current-voltage characteristics. One TER mechanism results from the dependence of the electrostatic tunneling barrier on the polarization orientation. We discuss a model allowing to estimate  this barrier modulation in the 2D limit. 

Moreover, a graphene-FE-graphene junction allows for a more sensitive finite voltage TER mechanism due to 2D momentum conservation. Resonance tunneling peaks in the $I(V)$ characteristics were identified both in planar 2D  junctions of semiconductor heterostructures~\cite{PhysRevB.44.6511,eisenstein1992coulomb}, and more recently in 2D materials~\cite{Britnell2013,mishchenko2014twist,ChenJulian2021}, specifically for graphene-h-BN-graphene~\cite{Britnell2013}. There, the h-BN material simply acts as a barrier~\cite{yan2021barrier,yan2022giant,yan2020significant}, whereas the resonance peaks emerge due to momentum conserving tunneling between the graphene electrodes.  The device we consider is almost the same as in Ref.~\onlinecite{Britnell2013}, where now the barrier consists of a parallel stacked h-BN bilayer supporting the polar AB or BA  interfaces. Such a device is expected to be a perfect candidate to employ resonance tunneling peaks as a sensitive probe of 
interfacial polarization, which essentially acts as an internal voltage that shifts the resonance. We indeed find using our self-consistent model a sizable shift of the resonance peak~\cite{Britnell2013,mishchenko2014twist,ChenJulian2021} for the two polarization directions. 

\begin{figure}
           \includegraphics[width=1.0\columnwidth]{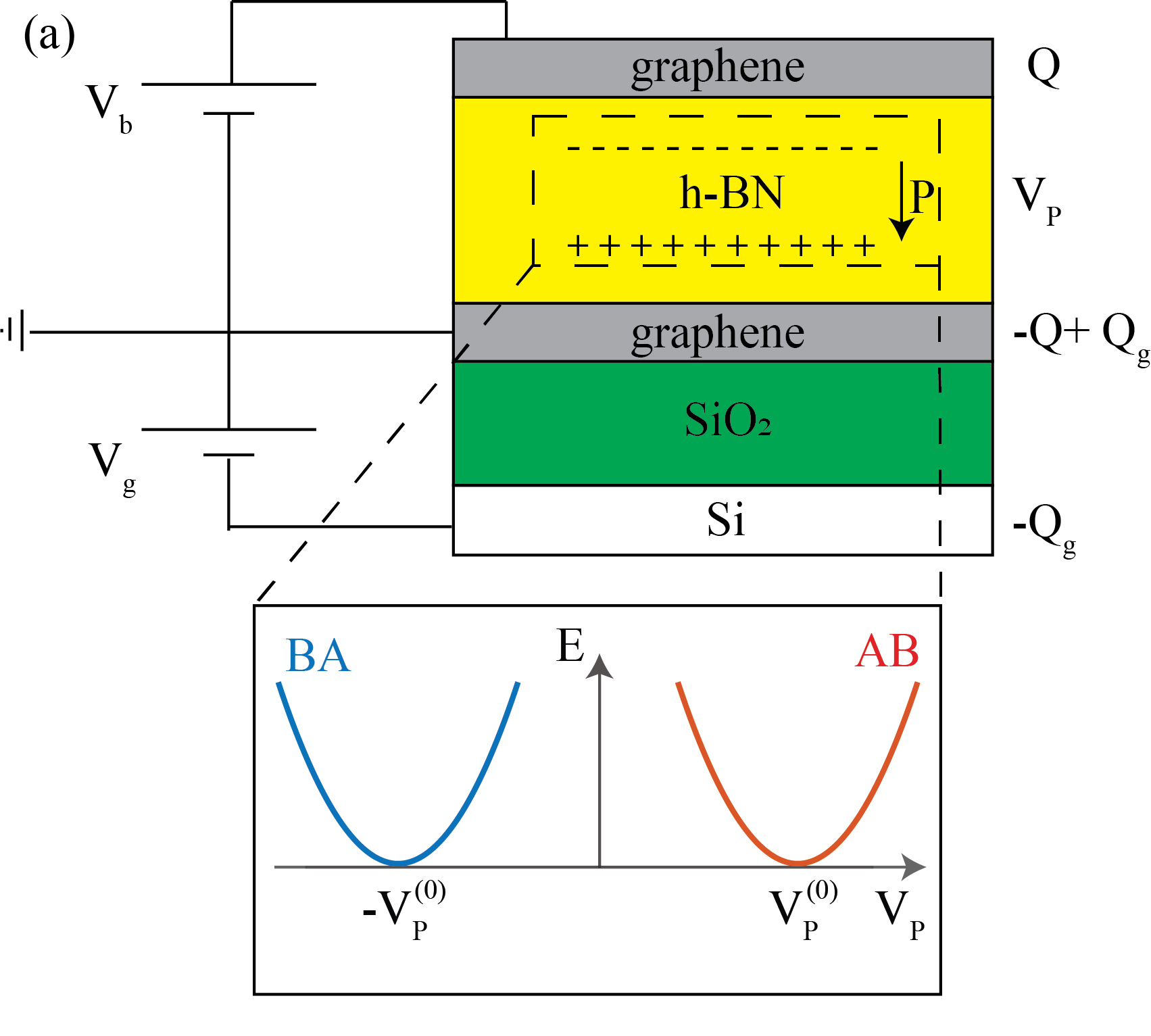}
         \caption{(a) FE tunnel junction: a pair of tunnel coupled graphene sheets and a bottom gate are modeled by a pair of plate capacitors $C_0,C_g$. The 
         polar bilayer interface is stacked inside the graphene-graphene capacitor and leads to an internal voltage denoted $V_{P}$. (b) Schematic 
         free energy of the 
         polarization of the interface, $V_{P}$. 
         The system consists of two degenerate minima corresponding to  stacking configurations  AB and BA. In our model Eq.~(\ref{eq:model}) the 
         polarization,  $V_{P}$, which takes the value $V_{P}^{(0)}$ for the bare bilayer,  interacts with the electric field due to the electrons in the graphene sheets and gate. \label{fig:device}
         }
 \end{figure}

The paper is organized as follows. After introducing the model in Sec.~\ref{se:model}, we study the Kelvin probe voltage in Sec.~\ref{se:kpvoltage}. In Sec.~\ref{se:TER} we estimate the ratio of the tunneling coefficients in our device. In Sec.~\ref{se:tunneling} we consider momentum conserving tunneling and focus on resonance peaks as a means to detect the polarization. We summarize in Sec.~\ref{se:summary}.
\section{Model
}
\label{se:model}

As shown in Fig.~1(a), we consider a tunneling junction  consisting of a 2D 
polar interface
encapsulated between two graphene sheets. Our theory is not restricted to a specific 
material, but for definiteness we consider parallel stacked bi-layer h-BN, having two polar stacking configurations denoted AB and BA exhibiting a finite 
out of plane 
polarization $P$ which is accounted for by an internal voltage denoted $V_{P}$, see Fig.~1(b). 
A tunneling current is enabled by a bias voltage $V_b$. 
The bottom graphene sheet is gated by a voltage $V_g$. The energy  describing the system  is given by
\bea
\label{eq:model}
    E(Q, Q_g,V_{P}) &=& \frac{Q^2}{2 C_0 } - Q (V_b + V_{P})  \nonumber \\
    &+& \frac{ Q_g^2}{2 C_g } -  Q_g V_g \nonumber \\
      &+& \frac{C_0}{2} \frac{\epsilon_r}{\epsilon_r-1} (V_{P} -p V_{P}^{(0)} )^2
    \nonumber \\
  &+&E_G (Q) +E_G (-Q+ Q_g).
\eea
Here $C_0$ is the capacitance between the two graphene sheets, and $C_g$ is the capacitance between the bottom graphene sheet and a gate. We measure energy, charge and capacitance per unit area.

The third line encapsulates a quadratic expansion of the free energy~\cite{ViznerStern2021} around the minima for either one of the two stacking configurations denoted by $p=\pm 1$. By construction, the bare KP voltage (i.e. without the electrodes) is given by $\pm V_{P}^{(0)}$ for AB or BA stacking, see Fig.~1(b). Its value $V_{P}^{(0)} \approx 120 {\rm{mV}}$ was measured directly  using Kelvin probe force microscopy~\cite{ViznerStern2021}. In addition, this term renormalizes the capacitance $C_0 \to C=C_0 \varepsilon_r$ by the dielectric constant of the material.

The last line represents the quantum capacitance of the graphene sheets, with
\begin{equation}
    E_G (Q) = \frac{2}{3} \beta |Q|^{\frac{3}{2}},  
\end{equation}
and $\beta = \sqrt{\frac{\pi}{e^3}} \hbar v_F$. $E_G (Q)$ is the total kinetic energy density measured from the Dirac point, valid in the vicinity of the Dirac point, and $Q>0(Q<0)$ refers to holes (electrons).

The goal of this electrostatic model is to allow variations of the polarization of the AB or BA interface due to electric fields produced by the graphene electrodes.

The equilibrium configuration 
is obtained by minimizing the energy with respect to the charge induced on the top sheet $Q$ and on the gate $Q_g$, and the polarization of the interface $V_{P}$
. The equation 
$\frac{\partial E}{\partial V_{P}} = 0 $ gives
\be
\label{eq:VFE}
V_{P} = p V_{P}^{(0)} + \frac{(\epsilon_r - 1)Q}{\epsilon_r C_0},
\ee
showing that the internal polarization is affected by the electrodes.  
Defining the Fermi energy 
\be
E_F(Q)=-e \frac{\partial E_G(Q)}{\partial Q}=-e~{\rm{sign}}(Q)\beta  |Q|^{1/2},~~~ (e>0),
\ee
the equations $\frac{\partial E}{\partial Q} = \frac{\partial E}{\partial Q_g} =0$ yield
\bea
\label{eq4}
       0 &=& \frac{Q}{C} - V_b -p V_{P}^{(0)} -E_F(Q)/e+E_F( Q_g-Q)/e,   \\
       \label{eq4a}
0 &=&      \frac{ Q_g}{C_g} -V_g - E_F( Q_g-Q)/e.
\eea
Eqs.~(\ref{eq4}) and (\ref{eq4a}) are subsequently 
solved numerically for $Q$ and $Q_g$ as function of $V_g$ and $V_b$ for either sign of the polarization $p=\pm 1$. 

\section{Kelvin probe voltage 
}
\label{se:kpvoltage}
\begin{figure*}
           \includegraphics[width=\textwidth]{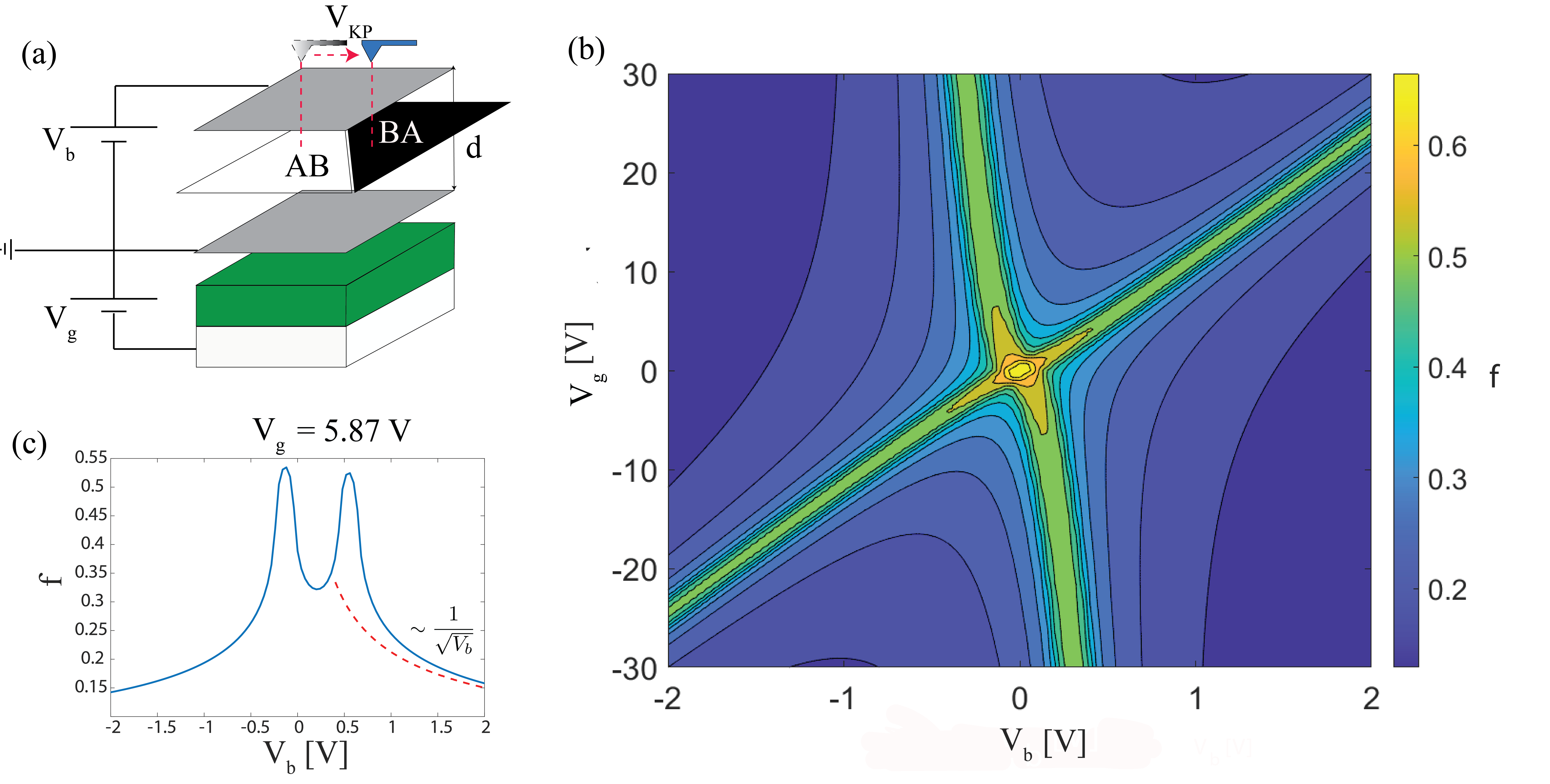}
         \caption{(a) Schematic view of the Kelvin probe measurement device. We define the Kelvin probe sensitivity (KPS) $f=\delta V_{KP}/(2 V_{P}^{(0)})$ [see Eq.~(\ref{eq:f})] as the relative difference in the voltage measured 
         for the two 
         polarization orientations of the two stacking configurations $AB$ and $BA$. (b) KPS versus gate and bias voltage. Maximal sensitivity occurs when one of the graphene sheets is at charge neutrality.
         (c) Cut along a fixed gate voltage showing a double peak. A fit (dashed line) to Eq.~(\ref{eq:sqrt}) shows that the KPS decays at large bias as $V_b^{-\frac{1}{2}}$. In this and all subsequent calculations we use $d = 5 {\rm{nm}}$, $d_g = 90 {\rm{nm}}$ (width of gate capacitor) and $\epsilon_r = 3.8$.
         }
 \end{figure*}

In this section we calculate the  KP voltage, which can be measured via scanning atomic force microscopy (AFM) as function of gate and bias voltages, in a setup as in Fig.~2(a). Such measurements were performed for the bare ferroelectric material~\cite{yasuda2021stacking,ViznerStern2021,weston2022interfacial}. Our theory addresses the renormalization of the internal polarization $V_{P}$ and the resulting total KP voltage $V_{KP}$ across the electrodes.

The total voltage drop measured in a Kelvin probe setup as in  Fig.~2(a) is given by
\begin{equation}\label{eq:kpvoltage}
    V_{KP}  \equiv V_{P} - \frac{Q}{C_0} =  p V_{P}^{(0)}- \frac{Q}{ C}, 
\end{equation} 
where we used Eq.~(\ref{eq:VFE}). We subsequently focus on the difference of $V_{KP}$ for the two polarization orientations, $\Delta V_{KP} =V_{KP}^{(+)}-V_{KP}^{(-)} $ which can be measured by scanning the AFM tip across a
stacking
domain wall, see Fig.~2(a). 

It is convenient to measure $\Delta V_{KP}$ in units of the bare polar domains voltage by defining the 
Kelvin probe sensitivity (KPS)
\be
\label{eq:f}
f \equiv \frac{\Delta V_{KP} }{2V_{P}^{(0)}}=1-\frac{Q^+-Q^-}{2C V_{P}^{(0)}}.
\ee
The second equality, obtained from Eq.~(\ref{eq:kpvoltage}), shows that $f \to 1$ in the absence of charges on the electrodes.

The numerical solution for $\Delta V_{KP}$ and hence the KPS is shown in Fig.~2(b) as function of bias and gate voltages. A cut with fixed $V_b$ in shown in Fig.~2(c). We can see that $\Delta V_{KP}$ peaks when either one of the two graphene sheets is at charge neutrality and takes a value which is of the order of $V_{P}^{(0)}$ (and hence $f \lesssim 1$). Along these peaks the charge transfer between the electrodes is small and primarily determined by quantum capacitance, and hence it very inefficiently screens electrostatically the internal polarization. 

We now discuss the physics describing the tail and peaks of $\Delta V_{KP}$ (or equivalently $f$ in Fig.~2(c)). The tails can be analysed by considering the quantum capacitance as a small perturbation. One can solve Eqs.~(\ref{eq4}) and (\ref{eq4a}) up to a given order in $\beta$. In the absence of quantum capacitance ($\beta \to 0$) we have 
\be
\label{eq:Q0}
Q \to Q^{(0)} \equiv C (V_b \pm V_{P}^{(0)}),~~~Q_g \to  Q_g^{(0)} \equiv C_g V_g.
\ee
Then $V_{KP}^\pm = V_b$, $\Delta V_{KP}=0$, and $f \to 0$. The vanishing KPS is expected since the potential on the electrodes adjusts such as to completely screen the internal polarization.
Considering a small quantum capacitance correction, $Q=Q^{(0)}+ \delta Q$, $Q_g=Q_g^{(0)}+ \delta Q_g$, and expanding Eq.~(\ref{eq4}) to linear order in these deviations, we obtain for $V_b \gg V_{P}^{(0)}$
\be
\label{eq:sqrt}
f \to  \beta \sqrt{\frac{C}{V_b}}.
\ee
This is confirmed as a dashed line in Fig.~2(c).

The height of the two peaks in Fig.~2(c) can be obtained by considering the quantum capacitance of the corresponding neutralized  graphene sheet as the dominant term. Consider for example the $Q=0$ peak [nearly vertical peak ridge in Fig.~2(b)]. We can then decouple Eqs.~(\ref{eq4}) and (\ref{eq4a}) by replacing the Fermi energy of the bottom layer $E_F(Q_g-Q)$ by $E_F(Q_g)$. We thus obtain
$0=\frac{Q}{C}-V_b-p V_{P}^{(0)} - E_F(Q)/e+E_F(Q_g)/e$, yielding a peak at $V_b=E_F(Q_g)/e$ where for large $V_g$ we  have $Q_g=C_g V_g$. Solving the quadratic equation yields $Q=p \left( -C \beta+\sqrt{(C \beta)^2 + 4 C V_{P}^{(0)}} \right)^2/4$. Substituting in Eq.~(\ref{eq:f}), we have $f=1-\frac{Q}{C V_{P}^{(0)}}$. Expanding in small $V_{P}^{(0)}$, yields
\bea
f=1-\frac{V_{P}^{(0)}}{C \beta^2}  +\mathcal{O}((V_{P}^{(0)})^2).
\eea
We can see that $C \beta^2 $ sets a voltage scale below which quantum capacitance effects set in. In our system 
\be
\mathcal{V} \equiv C \beta^2= \frac{e}{16 \pi d \varepsilon_0} \left( \frac{4 \pi \varepsilon_0 \hbar v_F}{e^2} \right)^2\sim 0.3V.
\ee
Thus, the reason that the peaks in KPS approach nearly unity 
is that the material property $V_{P}^{(0)}$ is small, but of the order of $\mathcal{V}$. 

\section{Polarization dependent tunneling coefficient}
\label{se:TER}
In this section we discuss 
the influence of the polarization on the tunneling coefficient $|\mathcal{T}|^2$. This effect can be understood 
from a polarization-dependent distortion of the electrostatic tunnel barrier~\cite{zhuravlev2005giant,gerra2007ferroelectricity}, modifying the  tunneling coefficient due to its exponential sensitivity $|\mathcal{T}|^2 \propto e^{-2 \kappa d}$. 

Let us denote the tunneling amplitudes for the two polarization orientations by $\mathcal{T}^\pm$ and define the relative barrier modulation
\be
\label{eq:RBM}
\eta=\frac{|\mathcal{T}^+|^2-|\mathcal{T}^-|^2}{|\mathcal{T}^+ |^2 +| \mathcal{T}^-|^2}.
\ee
As a simple model providing an order of magnitude estimate for $\eta$,
consider the bottom and top graphene sheets to be located at the $z=\pm d/2$ planes, and the two h-BN layers at the $z=\pm d/6$ planes. The tunneling amplitude from bottom to top  is given via 3-rd order perturbation theory by~\cite{PhysRevApplied.2.014003} $\mathcal{T} \propto \frac{t^3}{E(z=-d/6)E(z=d/6)}$, where $t$ are nearest layer hopping amplitudes, and $E(z=\pm d/6)$ is the energy in the h-BN layers. We denote by $E_g$ the energy gap of the bare h-BN layer at $Q=0$. Adding the linear potential due to the charged electrodes, $E(z)=E_{g}+\frac{e Q z}{dC}$.
The tunneling amplitude  becomes $\mathcal{T}^\pm \propto \left( E_g^2-\left( \frac{e Q^\pm }{6C} \right)^2 \right)^{-1}$. If the two directions of polarization lead  to exactly opposite charge transfer, $Q^+ = - Q^-$, then $\mathcal{T}^+=\mathcal{T}^-$. In general $\eta=0$ when  inversion symmetry holds, namely when $V_b=0$ and $V_g=0$. Finite
barrier modulation results from $Q^+ \ne - Q^-$,
 \be
\eta \approx \frac{\left( \frac{eQ^- }{6C} \right)^2-\left( \frac{eQ^+ }{6C} \right)^2}{2E^2_g}.
 \ee
For sufficiently large $V_b$, using Eq.~(\ref{eq:Q0}) which ignores quantum capacitance effects, we have $\eta= - e^2 V_b V_{P}^{(0)}/(18 E_g^2)$ independently of $V_g$. Assuming $E_g \sim 1eV$, this result leads to a 
few percent relative barrier modulation for a bias $V_b$ of few volts. The smallness of the effect derives from the small ratio between the voltage $V_{P}^{(0)}$ and the gap of h-BN, however this can be made larger in other materials.  Higher $\eta$ may also be obtained in an asymmetric device when one electrode is weakly coupled while the other one is strongly coupled~\cite{rogee2022ferroelectricity}.


We note that the TER is defined  [see Eq.~(\ref{eq:TERdef}] below and  Refs.~\onlinecite{tao2016ferroelectricity,kang2020giant,yan2022giant}) in terms of the currents at finite bias voltage. Our definition of the relative barrier modulation in Eq.~(\ref{eq:RBM}) allows to separate the effect of the modulation of the barrier itself, which leads to a relatively week contribution to TER, from the effects associated with momentum conserving tunneling between two graphene sheets, to be considered in the next section. 

\section{Momentum conserving Tunneling}
\label{se:tunneling}

In this section we consider 2D momentum conserving tunneling through the 
polar interface. In conventional tunnel junctions 2D momentum conservation leads to a resonance peak in the $I(V)$ characteristics corresponding to two aligned Dirac cones~\cite{Britnell2013,mishchenko2014twist,ChenJulian2021}. Our goal is to incorporate the interfacial
polarization into the resonant condition, yielding a sensitive probe of the polarization orientation and magnitude.

Following the model outlined in Ref.~\onlinecite{Britnell2013}, we start with a Fermi golden rule expression for the tunneling current
\begin{equation}
\begin{split}
    I^{(\pm)}  = |\mathcal{T}^\pm|^2 \sum_{\nu_B,\nu_T } & \int d^2 k_T \int d^2 k_B   \left( f_B(\varepsilon_{\vec{k}_B,\nu_B}) - f_T(\varepsilon_{\vec{k}_T,\nu_T}) \right)  \\  & (V_q(k_B,k_T))^2 \delta\left( \varepsilon_{\vec{k}_B,\nu_B}  -\varepsilon_{\vec{k}_T,\nu_T} -e V_{KP}  \right).
\end{split}
\end{equation}    
In this section, in order to disentangle the TER effects of the previous chapter with effects of momentum conservation, we assume for simplicity $\mathcal{T}^+ = \mathcal{T}^-$. Here, $f_{i}(\varepsilon)=(\exp[(\varepsilon-E_{F,i})/T]+1)^{-1}$ are Fermi functions and $\varepsilon_{\vec{k}_i,\nu_i} = \hbar v_F  \nu_i |\vec{k}_i|$ are the electron ($\nu=1$) and hole ($\nu=- 1$) bands  in the bottom or top  sheet $(i=B,T)$, respectively (see Fig.~3). This process describes tunneling of an electron from an occupied momentum state $\vec{k}_B$ in the bottom layer to an unoccupied state with momentum $\vec{k}_T$ at the top layer,  measured from the respective Dirac point in a given valley. The scattering potential 
contains a phenomenological momentum dependence  $V_q (k_B,k_T)= \frac{1}{q_c^2 + (\vec{k}_B - \vec{k}_T - \vec{Q} )^2} $. Here $|\vec{Q}|=K \theta $ is a momentum shift between the top and bottom Dirac points (at  momentum $K$ in the Brillouin zone) due to a relative twist of the layers by angle $\theta$.  The limit $q_c \to 0$ corresponds to momentum conserving tunneling. A finite $q_c$ phenomenologically describes non-momentum-conserving tunneling processes e.g. due to short-range disorder or the Moir\'e pattern of either the h-BN or its interface with graphene. We take $q_c^{-1}=12 nm$~\cite{Britnell2013}. The typical energy band diagram in the presence of a bias voltage is shown in Fig.~3(a). Since $E_{F,T}=E_F(Q)$ and $E_{F,B}=E_F(Q_g-Q)$ mark the distance of these Fermi levels from the corresponding Dirac point; from Eq.~(\ref{eq4}) it follows that the energy difference of the two Dirac points is given precisely by $eV_{KP}$, as marked in Fig.~3(a). 
\begin{figure}
         \includegraphics[width=1.0\columnwidth]{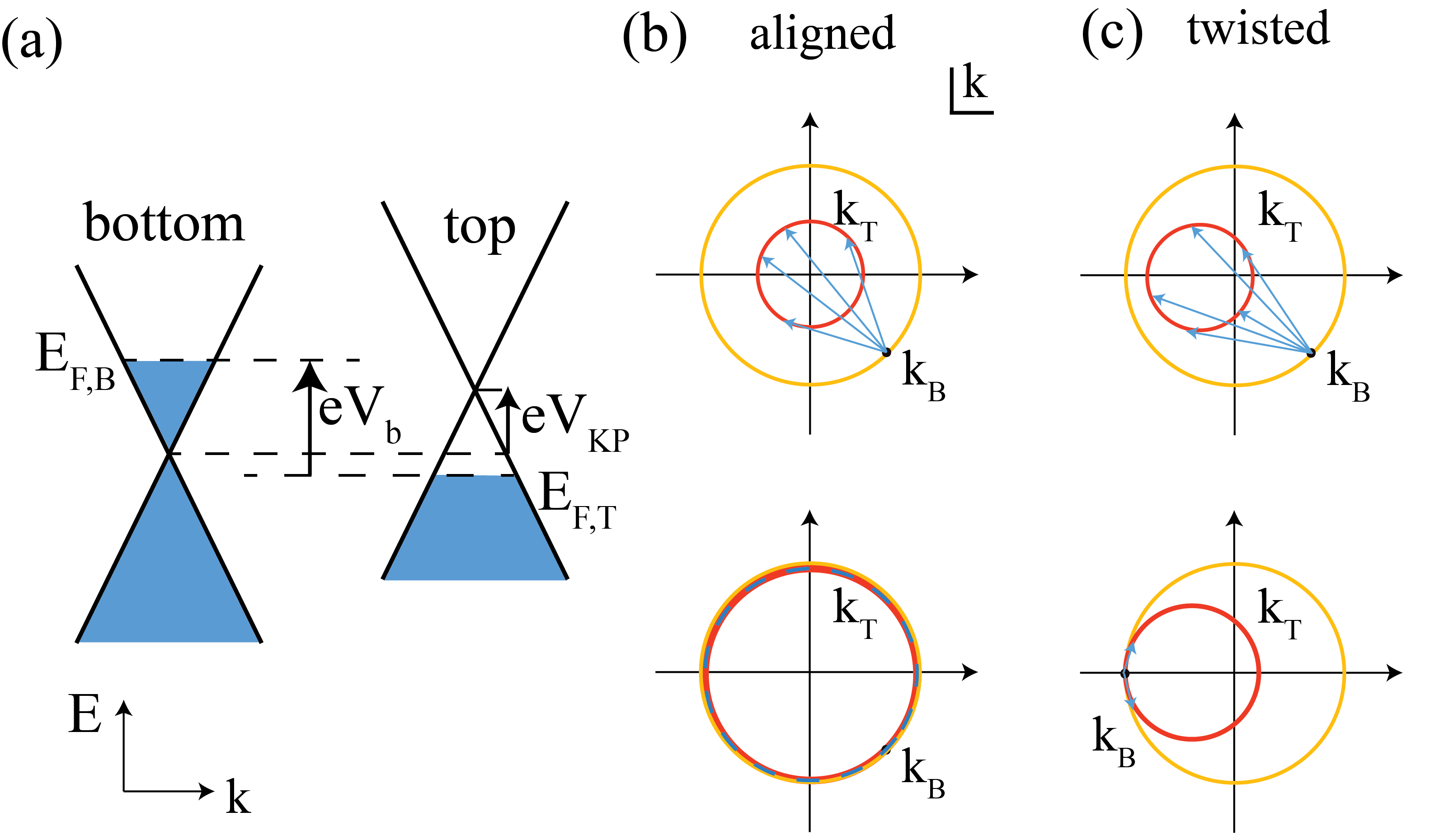}
         \caption{(a) Typical energy band diagram (momentum shift for clarity). The Fermi levels of the bottom and top layers $E_{F,i}$ ($i=B,T$) are determined from the charge densities $Q,Q_g$. The $KP$ voltage $V_{KP}$ sets the energy  misalignment of the Dirac points. (b) At zero relative twist angle equi-energy contours are concentric circles. At $V_{KP}=0$ these rings exactly overlap for all energies (see marked energy diagram in Fig.~4), leading to a resonant momentum conserving tunneling in the entire voltage window between the two Fermi energies.  (c) For finite twist the equi-energy circles are non-concentric, and resonance peaks occur when these circles are tangential for all energies (see marked energy diagrams in Fig.~5), when $eV_{KP} = \pm \Delta$ is satisfied (see Eq.~(\ref{eq:VKPDELTA})).\label{fig:3}
         }
 \end{figure}
 
 \subsection{Zero twist angle}
 When the two graphene sheets are perfectly aligned we have $\vec{Q}=0$. For any energy within the voltage window, the two momenta $\vec{k}_B$ and $\vec{k}_T$ belong to two concentric circles in momentum space, as denoted in Fig.~3(b). The energy displacement of the Dirac cones, $e V_{KP}$, is controlled by the bias voltage. When $V_{KP}=0$ these two circles in momentum space overlap, for all energies, leading to a resonant current peak. 

The resonance peak can be obtained  by performing the angular integration, leading to
\begin{equation}
\label{eq:32}
\begin{split}
    I^{(\pm)} \propto  \int_{-\infty}^\infty d k_T d k_B &  \frac{(f_B - f_T) |k_B| |k_T| (q_c^2 + k_B^2 + k_T^2) }{\left(q_c^2 + (k_B - k_T)^2 \right)^\frac{3}{2} \left(q_c^2 + (k_B + k_T)^2 \right)^\frac{3}{2}  } \\
    & \delta\left( \hbar v_F(k_B -  k_T) - e V_{KP}  \right).
\end{split}    
\end{equation}
This yields a single integral that we evaluate numerically.

\begin{figure*}
           \includegraphics[width=\textwidth]{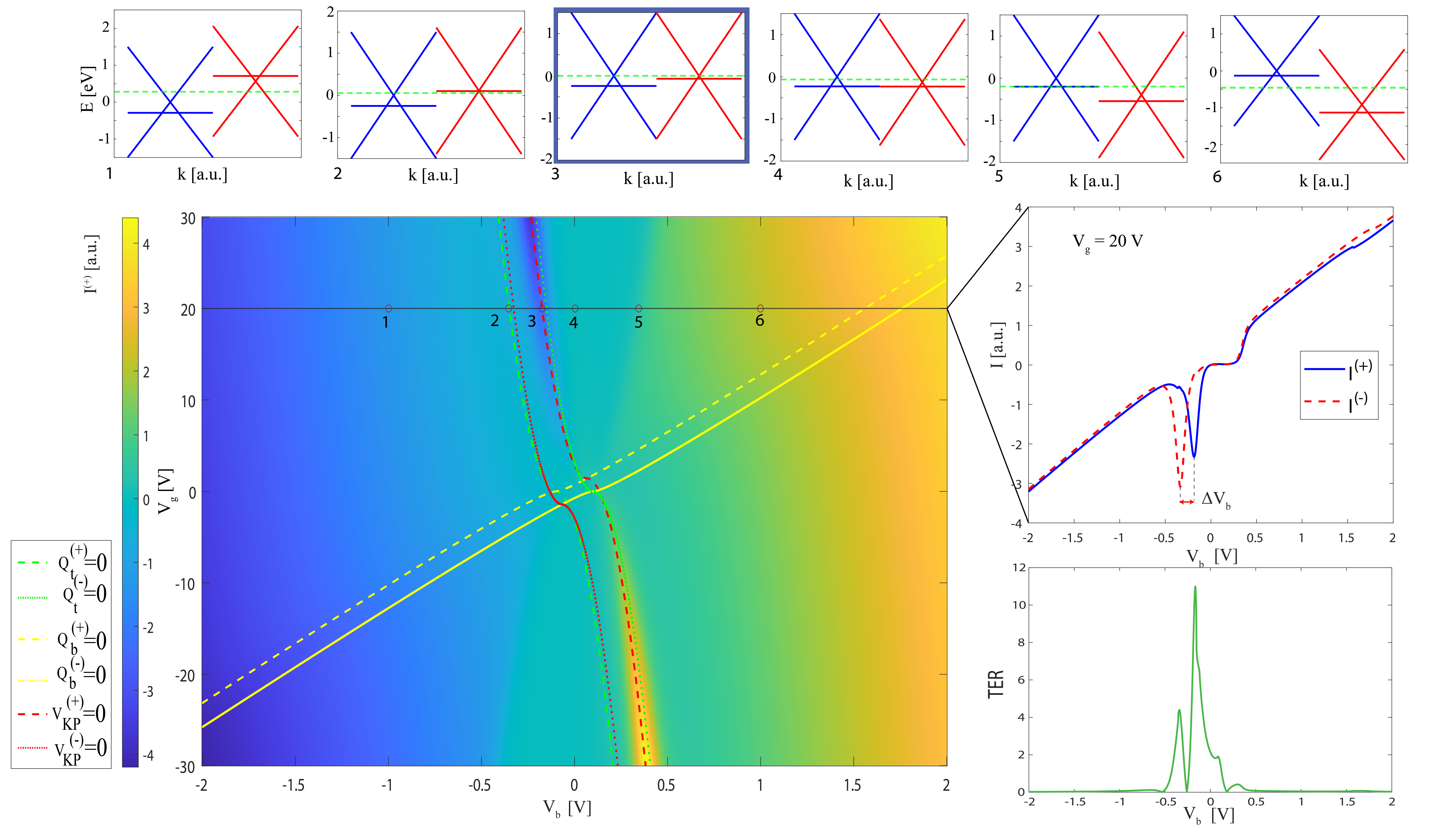}
         \caption{Color plot showcasing the calculated current $I^{(+)}$ for one orientation ($p = +1$) of the polarization, as a function of the gate and bias voltage, for the untwisted case $\theta=0$. On top of it are overlayed the curves for which the top and/or the bottom graphene electrodes are at the Dirac point (green $Q_{t}^{(p)}=0$ and yellow $Q_{b}^{(p)}=0$ curves, respectively), as well as the curves where the graphene sheets are electrostatically aligned (red curve, $V_{KP}^{(p)}=0$). 
         The right inset shows a cut of the $I-V$ curve at $V_g=20V$ for both $p=\pm 1$.
         The two resonance peaks match with the band alignment condition $V^{(\pm)}_{KP}=0$ and their line shape is given by Eq.~(\ref{eq:32}). The top inset displays the Dirac bands (horizontally shifted for clarity) for $p=1$, for points 1-6 corresponding to $V_g=20V$ and $V_b=-1,-0.35,-0.17,0,0.34,1.0$V. The resonance case corresponds to  point 3 (blue outline). The green dashed lines represent the energy midway between the two Dirac points, at which the tunneling process   conserves  momentum. Tunneling through this energy state begins at point 5, at which the  plateau in the right inset ends.
         }
 \end{figure*}

Fig.~4 shows a color map of the current $I$ versus $V_b$ and $V_g$ for a specific polarization direction.  As a guide to the eye, this figure displays the charge neutrality curves of either bottom or top graphene sheets.
We also plot the curve where $V_{KP}=0$. A cut of the $I(V_b)$ curve for either direction of the polarization for fixed $V_g$ is shown in the right-top inset of Fig.~4. We can see a pronounced resonance peak positioned precisely at $V_b$ at which $V_{KP}=0$ where the Dirac cones overlap. This occurs for different resonant voltages for the two polarization orientations.  

The line shape of the peak versus $V_b$ stems from the implicit dependence of $V_{KP}$ on the latter, and takes the form 
\be
\label{eq:32b}
I(V_b)|_{V_{KP} \approx 0} \propto ((\hbar v_F q_c)^2+(e V_{KP})^2)^{-3/2}.
\ee

We can also observe a plateau in the $I(V_b)$ curve in Fig.~4 which terminates at $V_b \approx 0.4V$ (at point 5). As shown in the energy diagrams in  Fig.~4, this corresponds to a threshold for transport through the midway energy between the two Dirac points, see green dashed lines. At the threshold $V_b \approx 0.4V$ this specific energy enters into the voltage window~\cite{feenstra2012single}.

We note that the shifted resonance peaks lead to a significant TER~\cite{wu2020high,yan2021barrier,yan2022giant}, defined as
\be
\label{eq:TERdef}
{\rm{TER}} = \frac{|I^{(+)} - I^{(-)} |}{\min(|I^{(+)}|,|I^{(-)}|)},
\ee
as shown in Fig.~4 (right-bottom inset). The role of our definition of the relative barrier modulation $\eta$ in Sec.~\ref{se:TER} meant to separate this effect from the modification of the tunneling amplitudes for the two polarization directions.
\subsection{Finite twist angle}
Tunneling between  twisted graphene sheets has been discussed in numerous theory~\cite{feenstra2012single,de2014theory} and experimental~\cite{mishchenko2014twist} works. Whereas at zero twist, there is a single resonance condition $V_{KP}=0$ at which the two Dirac cones completely overlap, at a finite twist angle there is a momentum shift between the Dirac points, given by $|\vec{Q}|= K \theta$ for $\theta \ll 2 \pi$. Thus equi-energy contours of the bottom and top layers are non-concentric circles, see Fig.~3(c). Upon tuning their relative energy $e V_{KP}$ there are two situations where the Dirac cones are tangential, 
\be
\label{eq:VKPDELTA}
e V_{KP} = \pm \hbar v_F \theta K \equiv \pm \Delta.
\ee

The calculated current with a finite twist angle of $\theta = 3.5^{o}$ is shown in the color plot of Fig.~5. We can clearly see that the pair of resonance peaks, also shown along a cut of fixed gate, overlaps with the condition Eq.~(\ref{eq:VKPDELTA}). The corresponding dispersion relations are shown in the top insets of Fig.~5. The sensitivity of the resonant peaks to different orientation of the polarization can be enhanced by tuning the gate such that the resonance peaks will occur near the charge neutrality condition of one graphene sheet. This is indeed seen for the cut of the $I(V_b)$ curve at $V_g=9.5V$.

\begin{figure*}
           \includegraphics[width=\textwidth]{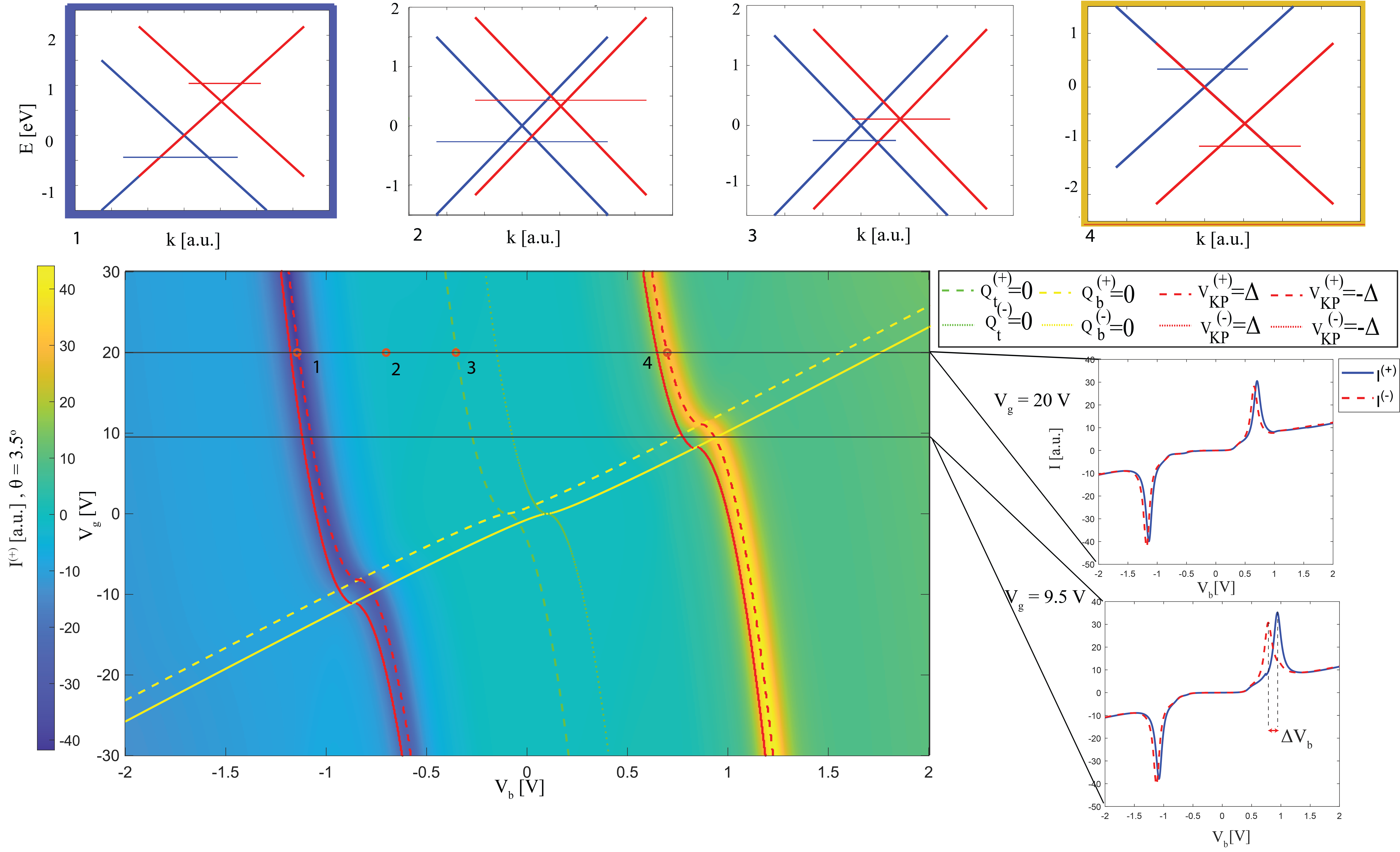}
         \caption{Color plot showcasing the calculated current as a function of gate and bias voltage for specific orientation of the polarization $(p=+1)$, with twist angle $\theta = 3.5^{o}$ between the graphene sheets. The resonance peak splitting to two peaks at $eV_{KP}= \pm \Delta$ is observed on the color plot. The right insets show two cuts at fixed gate voltages, plotting both $I^{(+)}$ and $I^{(-)}$. We can see that each peak of either $eV_{KP}= \Delta$ or $eV_{KP}= - \Delta$ further splits according to the direction of the polarization $(p=\pm 1)$. The top inset shows the calculated band structure alignment (for points 1-4 corresponding to $V_g=20V$ and $V_b=-1.14,-0.70,-0.35,0.70 $V). 
         }
 \end{figure*}
 
 

\subsection{Dependence on G-G twist angle}
We now discuss the twist-angle dependence of the resonance peaks. At $\theta=0$ we observed a resonance peak whose position $V_{b,{\rm{peak}}}^{(p)}$ depends on the polarization orientation $p=\pm $. The peak separation $\Delta V_b = |V_{b,{\rm{peak}}}^{(+)}-V_{b,{\rm{peak}}}^{(-)}|$ is a sensitive probe of the polarization, see Figs.~4 and 5. 

The peak separation is plotted in Fig.~6 as a function of $V_g$ and $\theta$. For zero twist angle we have seen in the inset of Fig.~4 that $\Delta V_b \sim 0.15 V$ for $V_g=20V$. We can see in Fig.~6 that at $\theta=0$ this significant peak separation persists for any value of $V_g$, and peaks near $V_g=0$. But at finite twist angle, the peak separation becomes visible only for specific values of $V_g$. For example for $\theta=3.5^o$, as we have seen in the inset of Fig.~5,  the peak separation is resolved only near $V_g \sim \pm 10V$. 
As discussed, this maximal sensitivity occurs when both Eq.~(\ref{eq:VKPDELTA}) holds and one graphene sheet is at charge neutrality $Q_{b}=0$. Therefore, in order to achieve good sensitivity in terms of resonance peak separation,  one should operate such a device at specific twist angle dependent value of the gate voltage. 

The slope of the high sensitivity spokes in the $(V_g,\theta)$ plane can be obtained analytically. By solving Eqs.~(\ref{eq4}) and (\ref{eq4a}), with the added condition of Eq.~(\ref{eq:VKPDELTA}), we obtain the expression for the bias voltage at which the peak occurs as a function of $V_g$ and $\theta$. For a given twist angle, the difference of such bias voltage for the two orientations of the polarization, representing the peak separation, is extremized analytically with respect to $V_g$. For large enough gate voltages (outside the range where both graphene sheets are near the charge neutrality), we find
\be
V_g \simeq \pm \frac{C}{C_g} \frac{\hbar v_F}{e} K \theta,
\ee
which perfectly fits the spokes in Fig.~6 (not shown). 

\begin{figure}[ht]
           \includegraphics[width=1.0\columnwidth]{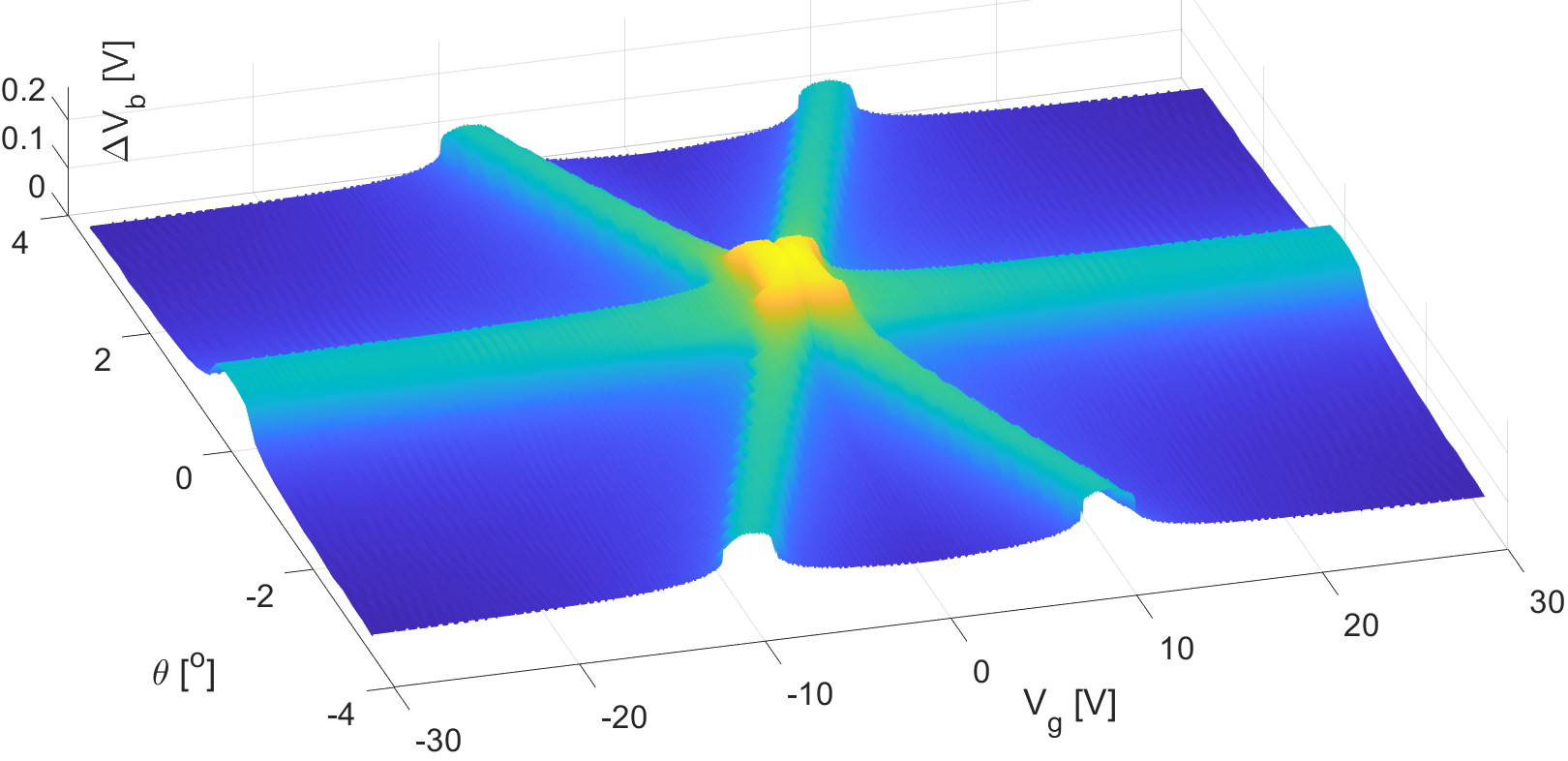}
         \caption{Peak separation $\Delta V_b$ as function of gate voltage and twist angle $\theta$. The peak separation is defined from the $I(V)$ characteristics as in the bottom inset of Fig.~5. For each $V_g$ and $\theta$, $\Delta V_b$ is the largest among the two peak separations. \label{fig:5}
         }
 \end{figure}
 
 \section{Summary}
To summarize, in this work we studied graphene-FE-graphene junctions. We focused on the self-consistent interplay of the polarization with the electrostatics and quantum capacitance of the graphene electrodes. We first asked whether experiments will detect a polar domain voltage across the conducting graphene electrodes, and discovered that a finite signal appears only upon gate tuning of one of the Dirac electrodes to charge neutrality. 

Moving to the tunneling current, we studied two mechanisms for its dependence  on the polarization direction. First, we discussed a dependence of the tunneling coefficient  on the polarization. This barrier sensitivity becomes sizable upon increasing the bias voltage which leads to a device asymmetry, but is relatively small at zero bias for available values of gate voltages. We then considered 2D momentum conserving tunneling, and showed that one can detect the actual value of the polarization via a shift in the voltage axis of a resonant peak occurring in the $I(V)$ characteristics due to the alignment of the Dirac cones. This resulted in a sizable TER.


In order to switch from one state to the other, one would need, in practice, to slide a domain wall separating AB and BA stacking through the tunneling region. This could be achieved by tuning the bias $V_b$ above the switching point. This process goes beyond the presented model. 

Our results apply to Moiré superlattice materials such as transition metal dichalcogenides. We hope that our theory will serve as a basis for analysis of near-future FE tunnneling experiments.

 \label{se:summary}
\section{Acknowledgments} 
We thank discussions with Moshe Ben Shalom, Igor Rozhansky, Simon Sallah Atri and Hadar Steinberg. This project received funding from the European Research Council (ERC) under the European Union’s Horizon 2020 research and innovation program under grant agreement No 951541. 

\bibliography{two_capacitor_lit.bib}

\widetext
\clearpage
\newpage
\setcounter{equation}{0}
\setcounter{figure}{0}
\setcounter{table}{0}
\setcounter{page}{1}
\renewcommand{\theequation}{S\arabic{equation}}
\renewcommand{\thefigure}{S\arabic{figure}}
\makeatletter
\renewcommand{\present@bibnote}[2]{}
\makeatother

\end{document}